\documentclass[prl,twocolumn,amsmath,amssymb,groupedaddress,superscriptaddress]{revtex4}
\usepackage{graphics}

\newcommand{\braket}[1]{\ensuremath{\left\langle{#1}\right\rangle}}
\newcommand{\ket}[1]{\ensuremath{\left|{#1}\right\rangle}}
\newcommand{\ad}{\ensuremath{a^\dagger}}
\def\be{\begin{equation}}
  \def\ee{\end{equation}}
\newcommand{\bd}{\ensuremath{b^\dagger}}

\begin{document}

\title{Fast and robust two--qubit gates for scalable ion trap
quantum computing}

\author{J.~J. \surname{Garc\'{\i}a-Ripoll}}
\email{Juan.Ripoll@mpq.mpg.de}
\affiliation{Max-Planck-Institut f\"ur Quantenoptik,
  Hans-Kopfermann-Str. 1, Garching, D-85748, Germany.}
\author{P. \surname{Zoller}}
\affiliation{Innsbruck Universit\"at, Technikerstr. 24, 6020
Innsbruck, Austria.}
\author{J.~I. \surname{Cirac}}
\affiliation{Max-Planck-Institut f\"ur Quantenoptik,
  Hans-Kopfermann-Str. 1, Garching, D-85748, Germany.}

\begin{abstract}
  We propose a new concept for a two--qubit gate operating
  on a pair of trapped ions based on laser coherent control techniques.
  The  gate is insensitive to the temperature of the ions, works also
  outside the Lamb--Dicke regime, requires no individual addressing by lasers,
  and can be orders of magnitude faster than the trap period.
\end{abstract}

\date{\today}

\maketitle

Trapped ions constitute one of the most promising systems to
implement scalable quantum computation. \cite{PhysicsTodayMay2003}
In an ion trap quantum computer qubits are stored in long-lived
internal atomic states. A universal set of single and two qubit
gates is obtained by manipulating the internal states with lasers,
and entangling the ions via the motional states
\cite{CiracZoller1}. During the last years a remarkable
experimental progress in building an ion trap quantum computer has
allowed to realize two--qubit gates
\cite{Wineland95,Wineland1,Wineland2,Blatt} and also to prepare
entangled states \cite{Wineland3,Wineland4,Wineland5}. The
ultimate challenge is now the development of scalable ion trap
quantum computing.  Scalability is based on storing a set of ions,
and moving ions independently, in particular to bring together
{\em pairs of
  ions} to perform a two-qubit gate \cite{Scalable-a,Scalable-b}.
Basic steps towards this goal have already been demonstrated
experimentally \cite{papermotionWineland}.

An important question to be addressed is to identify the current
limitations of the two--qubit gates with trapped ions (given the
fact that one--qubit gates are significantly simpler with those
systems). The ideal scheme should: (i) be independent of
temperature (so that one does not need to cool the ions to their
ground state after they are moved to or from their storage area);
(ii) require no addressability (to allow the ions to be as close
as possible during the gate so as to strengthen their
interaction), and (iii) be fast (in order to minimize the effects
of decoherence during the gate, and to speed up the computation).
This last property has been identified \cite{PhysicsTodayMay2003}
as a key limitation: in essentially all schemes suggested so far
\cite{CiracZoller1,Proposals-b,Proposals-c,Proposals-d,ProposalsMilburn,
  Proposals-a} one has to resolve spectroscopically the motional
sidebands of the ions with the exciting laser, which limits the laser
intensity and therefore the gate time.

\begin{figure}[t]
\begin{center}
  \resizebox{\linewidth}{!}{\includegraphics{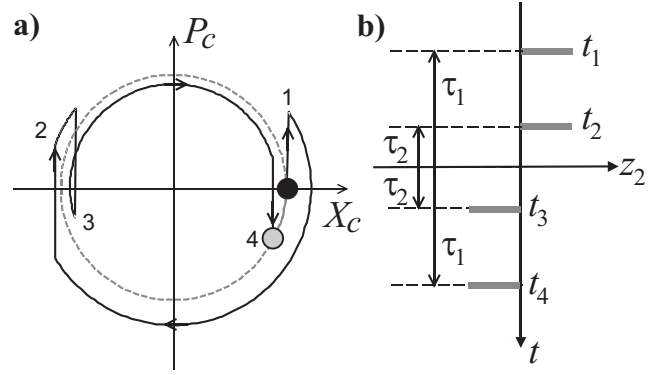}}
\end{center}
\caption{\label{fig-1} a) Trajectory in phase space of the
center-of-mass state of the ion $(X_c,P_c)$ (where
$(X_c+iP_c)/\sqrt{2}=\langle a \rangle$) during the 2-qubit gate
(solid line), connecting the initial state (black filled circle)
to the final state (grey filled circle) at the gate time $T$. The
time evolution consists of a sequence of kicks (vertical
displacements), which are interspersed with free harmonic
oscillator evolution (motion along the arcs). A pulse sequence
satisfying the commensurability condition (\ref{cond}) guarantees
that the final phase space point is restored to the one
corresponding to a free harmonic evolution (dashed circle). The
particular pulse sequence plotted corresponds to a four pulse
sequence given in the text (Protocol I). Figure b) shows how the
laser pulses (bars) distribute in time for this scheme.}
\end{figure}

The two--qubit gate between pairs of ions analyzed below solves
the problem of speed by using mechanical effects instead of
spectral methods to couple the motion and internal states of the
ions. In this way the new limits on the time of the quantum gate
are those of laser control, which can be orders of magnitudes
faster than the present limits dictated by trap design. Thus, the
method proposed here is a significant step forward towards fast
and efficient scalable quantum computations with trapped ions.

Below we will first study the dynamics of a pair of trapped ions
under the influence of short laser pulses with varying directions.
We will prove that there exist certain laser pulse sequences which
perform a phase gate on the two qubits, while leaving the motional
state unchanged. We illustrate this with two protocols for laser
pulses: (i) a sequence of four pulses which gives a gate time of
$T=1.08/\nu$ with $\nu$ the trap frequency, and (ii) a protocol
which allows us to perform a gate in a time $T\sim N_P^{-2/3}/\nu$
where $N_P$ is the number of laser pulses. Finally, we will
complement our study of the gate dynamics with an analysis of
possible errors, which includes fluctuations of the intensity or
the duration of the pulses, and temperature. The gate will be
shown to be extremely robust to these perturbations.

We consider two ions in a one--dimensional harmonic trap,
interacting with a laser beam on resonance. The Hamiltonian
describing this situation\cite{CiracZoller1} can be written as
$H=H_0 + H_1$, where $H_0 = \nu_c a^\dagger a + \nu_r b^\dagger b$
describes the motion in the trap and
\begin{eqnarray}
\label{hamil}
  H_1 &=& \frac{\Omega(t)}{2}
  \sigma_1^+ e^{i\eta_c (a^\dagger + a) + \tfrac{1}{2} \eta_r
    (b^\dagger +b)} + \nonumber\\
  && \frac{\Omega(t)}{2} \sigma_2^+ e^{i\eta_c (a^\dagger + a) - \tfrac{1}{2} \eta_r
    (b^\dagger +b)}  + h.c.
\end{eqnarray}
Here, $\nu_c=\nu$ and $\nu_r = \sqrt{3}\nu_c$ are the frequencies
of the center of mass and stretching mode, respectively; $a$ and
$b$ are the corresponding annihilation operators, and
$\eta_c=\eta/\sqrt{2}$ and $\eta_r=\eta\sqrt[4]{4/3}$ are
proportional to the Lamb--Dicke parameter, $\eta$. Note that the
Rabi frequency $\Omega$ is the same for both ions, since {\em we
have not assumed individual addressing}. Also notice that
replacing $\eta$ with $(-\eta)$ is equivalent to reversing the
direction of the laser beam.

In the following we will consider two different kind of processes:
(i) Free evolution, where the laser is switched off ($\Omega=0$)
for a certain time; (ii) Sequences of pairs of very fast laser
pulses, each of them coming from opposite sides. If we denote by
$\delta t$ the duration of a pulse and by $\Omega$ the
corresponding Rabi frequency, we are interested in the limit
$\delta t\to 0$ with $\Omega\,\delta t=\pi$. Processes (i) and
(ii) will be alternated (See Fig.~\ref{fig-1}a): at time $t_1$ a
sequence of $z_1$ pulses is applied, followed by free evolution
until at time $t_2$ another sequence of $z_2$ pulses is applied
followed by free evolution and so on. The $z_k$ are integer
numbers, whose sign indicates the direction of the laser pulses.

For a pulse sequence, consisting of kicks interspersed with free
harmonic time evolution (Fig.~\ref{fig-1}), we write the evolution
operator as ${\cal U}={\cal U}_c\, {\cal U}_r$, where ${\cal U}_{c,r}=
\prod_{k=1}^N U_{c,r} (\Delta t_k,z_k)$ has contributions of the
center--of--mass and relative motions,
\begin{subequations}
  \begin{eqnarray*}
    U_c(t_k,z_k) &=& e^{-i 2z_k\eta_c(a+a^\dagger)(\sigma_1^z+
      \sigma_2^z)} e^{-i\nu_c \Delta t_k a^\dagger a},\\
    U_r(t_k,z_k) &=& e^{-i z_k\eta_r(b+b^\dagger)(\sigma_1^z-
      \sigma_2^z)} e^{-i\nu_r \Delta t_k b^\dagger b}.
  \end{eqnarray*}
\end{subequations}
The integers $z_k$ indicate the direction of the initial pulse in
the sequence of pairs of very fast laser pulses, each of them
coming from opposite sites.

In order to fully characterize $U$, we only have to investigate
its action on states of the form $|i\rangle_1|j\rangle_2
|\alpha\rangle_c|\beta\rangle_r$, where $i,j=0,1$ denote the
computational basis, and $|\alpha\rangle$ and $|\beta\rangle$ are
coherent states. This task can be easily carried out once we know
the action of ${\cal U}=\prod_{k=1}^N U(\phi_k,p_k)$ on an
arbitrary coherent state $|\alpha\rangle$, where
\begin{equation*}
  U(\phi_k,p_k) = e^{-i p(a+a^\dagger)} e^{-i\phi_k a^\dagger a}.
\end{equation*}
We obtain ${\cal U}|\alpha\rangle = e^{i\xi} |\tilde
\alpha\rangle$, where
\begin{eqnarray*}
  \tilde \alpha &=& \alpha e^{-i\theta_N} - i \sum_{k=1}^N
  p_k  e^{i(\theta_k-\theta_N)},\\
  \xi &=& - \sum_{m=2}^N\sum_{k=1}^{m-1} p_mp_k
  \sin(\theta_k-\theta_m) - \Re \left[\alpha \sum_{k=0}^N p_k e^{-i\theta_m}
  \right],\nonumber
\end{eqnarray*}
with $\theta_k=\sum_{m=1}^k \phi_m$.

The crucial point is to realize that if $\sum_{k=1}^N p_k
e^{i\theta_k}=0$ the motional state $|\alpha\rangle$ after the
evolution is the same as if there was only free evolution
[Fig.~\ref{fig-1}a], and a global phase $\xi$ appears which does
not depend on the motional state. Translating this result to the
operators ${\cal U}_c|\alpha\rangle$ and ${\cal
U}_r|\beta\rangle$, we obtain the conditions
\begin{subequations}
\label{cond}
\begin{eqnarray}
  C_c&\equiv&\sum_{k=1}^N z_k e^{-i \nu t_k}=0,\\
  C_r&\equiv& \sum_{k=1}^N z_k e^{-i \sqrt{3}
    \nu t_k}=0.
\end{eqnarray}
\end{subequations}

If these commensurability conditions are satisfied, the motional
state of the ion will not depend on the qubits and the evolution
operator will be given by
\begin{equation} \label{evolve}
{\cal U}(\Theta)= e^{i\Theta\sigma_1^z\sigma_2^z}e^{-i\nu_cT\ad a}
e^{-i\nu_r T\bd b}.
\end{equation}
The value $T$ is the total time required by the gate and
\begin{equation}
  \label{phase}
  \Theta = 4\eta^2 \sum_{m=2}^N\sum_{k=1}^{m-1} z_k z_m
  \left[\frac{\sin[\sqrt{3}\nu\Delta t_{km}]}{\sqrt{3}}-\sin(\nu\Delta
  t_{km})\right],
\end{equation}
where $\Delta t_{km}=t_k-t_m$. Therefore, if Eqs. (\ref{cond}) are
fulfilled, and $\Theta=\pi/4$ we will have a controlled--phase
gate (which is equivalent to a controlled--NOT gate up to local
operations) which is {\em completely independent of the initial
motional state}, i.e. there are no temperature requirements.

\begin{figure}
  \centering
  \resizebox{\linewidth}{!}{\includegraphics{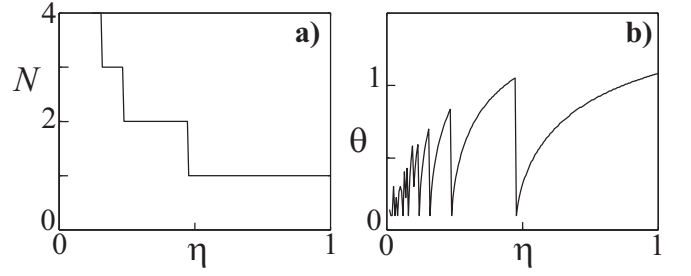}}
  \caption{\label{fig-2}
    (a) Number of pairs of pulses , and (b) relative angle of the
    two laser beams required to produce a phase gate using the
    first exact scheme developed in the paper.}
\end{figure}

It is straightforward to show that for any value of the time $T$
it is always possible to find a sequence of laser pulses which
implements the gate, and therefore the gate operation can be, in
principle, arbitrarily fast. The search for a sequence of pulses
may be done numerically, or even semianalytically. In the
following we give two simple protocols which are not optimized in
order to reduce the number of pulses.

The first protocol (Protocol I) requires the least number of
pulses to produce the gate in a fixed time $T\simeq
1.08(2\pi/\nu)$. The recipe is illustrated in Fig.~\ref{fig-1},
which provides the phase space plots for the evolution of the
motional state. The sequence of pulses is defined as $(z_n/N,
t_n)= \{(\gamma,-\tau_1), (1,-\tau_2), (-1,\tau_2),
(-\gamma,\tau_1)\}$. Here $0 < \gamma = \cos(\theta) < 1.0$ is a
real number, which may be introduced by tilting both lasers a
small angle $\theta$ with respect to the axis of the trap, so that
no transverse motion is excited. It is always possible to find a
solution to Eq.~(\ref{cond}) with $\tau_1 \simeq
0.538(4)(2\pi/\nu) > \tau_2 > 0$. The results for the performance
of the gate are summarized in Fig.~\ref{fig-2}. As shown in
Fig.~\ref{fig-2}(a), for realistic values of the Lamb--Dicke
parameter \cite{Wineland1} we only need to apply the sequence of
pulses one or two times to implement a phase gate.

The second protocol (Protocol II) performs the gate in an
arbitrarily short time $T$. The pulses are now distributed
according to $(z_n/N,t_n)= \{(-2,-\tau_1), (3,-\tau_2),
(-2,-\tau_3), (2,\tau_3), (-3,\tau_2), (2,\tau_1)\}$. The whole
process takes a time $T=2\tau_1$ and requires $N_p=\sum |z_n|=14N$
pairs of pulses. As Fig.~\ref{fig-2} shows, the number of pulses
increases with decreasing time as $N_p\propto T^{-3/2}$.

\begin{figure}
  \centering
  \resizebox{\linewidth}{!}{\includegraphics{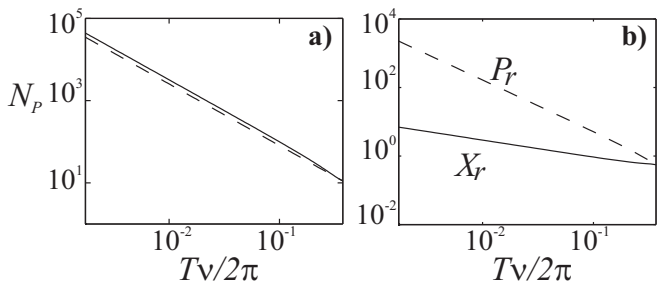}}
  \caption{\label{fig-3}
    (a) Log-log plot of the number of pairs of pulses required to
    produce a phase gate using protocol II, as a function of the
    duration of the gate, $T$, for a realistic value\cite{Wineland1}
    of the Lamb--Dicke parameter, $\eta=0.178$. We plot both the exact
    result (solid line) and a rough estimate $N_P = 40 (\nu
    T/2\pi)^{-3/2}$ (dashed line) based on perturbative calculations.
    (b) Maximum relative displacement, $X_r$ (solid), and maximum
    momentum acquired, $P_r$ (dashed line), for scheme II. These
    quantities are adimensionalized versions of the real observables,
    $X_r=\max[\braket{x_r(t)}/a_0]$, and
    $P_r\max[\braket{p_r(t)}a_0/\hbar]$.}
\end{figure}

In order to study the main potential limitations of our scheme, we
define the error of the gate $E$ in terms of the gate fidelity
\cite{Nielsen} as $ E = 1- \mathrm{Tr}_{\rm mot}\left\{Q_{\rm mot}
  \rho_{\rm mot} Q_{\rm mot}^\dagger\right\}$. Here
$\mathrm{Tr}_{\rm mot}$ and $\mathrm{Tr}_{\rm int}$ denote traces
over motional and internal degrees of freedom, and $ Q_{\rm mot} =
\mathrm{Tr}_{\rm int}\left\{{\cal U}(\pi/4) U_{\rm
real}^\dagger\right\}, $ depends on $U_{\rm real}$, the gate
performed in the presence of imperfections.

We now turn to a discussion of the possible sources of errors. A
limiting factor for the gate is the anharmonicities of the
restoring forces. The more pulses we apply, the larger the
relative displacement of the ions, as Fig.~\ref{fig-3}(b) shows.
When the ions become too close to each other, the increasing
intensity of the Coulomb force can lead to a breakdown of the
harmonic approximation which is implicit in Eq.~(\ref{hamil}). In
order to analyze this effect, we have made a perturbation analysis
for $\nu T\ll 1$ and found that such an anharmonicity causes an
error $E\simeq |0.4 a_0/d|^2/(2\pi\nu T)$, where $a_0$ is the
ground state size of the external potential and $d$ is the ion
separation in equilibrium. For typical parameters and imposing an
error $E\simeq 10^{-4}$ we obtain $\nu T\simeq 10^{-3}$. A similar
analysis could be applied to study anharmonicities of the trap
itself.

\begin{figure}
  \centering
  \resizebox{\linewidth}{!}{%
    \includegraphics{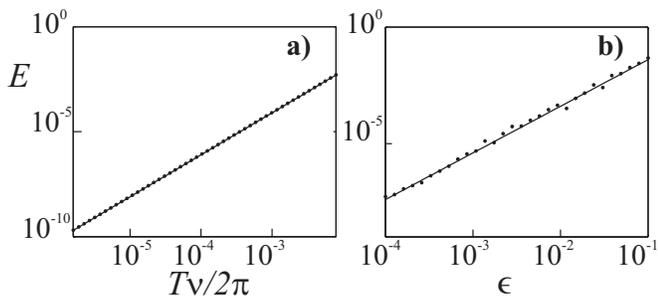}}
  \caption{\label{fig-4}
    For the sequences of four kicks designed in the paper,
    using\cite{Wineland1} $\eta=0.178$ and a duration of the gate
    $T={\cal O}(1/\nu)$, we have computed the dependence of the gate error
    with respect to the characteristics of the laser pulses. In
    (a) the error is plotted versus the duration of the pulse, $\tau$,
    rescaled using the period of the trap. It perfectly fits the
    estimate $E=2(\tau\nu/2\pi)^2$.  In (b) the mean error of the gate is
    plotted against the amplitude of random errors in the duration of the
    laser pulses. Our errors are modeled setting $\tau_k = \pi /(2\Omega) (1 +
    \epsilon r_k)$, with random numbers $r_k$ uniformly distributed in
    $[-1/2,1/2]$. In solid lines we show a visual aid which follows the
    formula $E=4\epsilon^2$.    }
\end{figure}

In addition, we have studied the influence of errors in the laser
pulses of our scheme.  Up to now, our analytical calculations
assumed that the intensity of the laser is very large during each
pulse, and that therefore one may neglect the influence of the
trap during this process. To validate this assumption we have
simulated numerically a system of two ions with only one
vibrational mode. We have used the exact sequences developed above
to produce the phase gate using only eight laser pulses. In Fig.~
\ref{fig-4}(a) we plot the error of the gate as a function of the
duration of the laser pulse, $\tau=\pi/2\Omega$. The longer the
pulse, the more important the effect of the trap, and the larger
the error. But even for relatively long pulses, we obtain a
fidelity which is comparable to the results obtained in current
setups \cite{Wineland1,Blatt,Wineland2}.  We have also studied the
influence of noise in the intensity of the laser pulses, or, what
is equivalent, random errors in its duration.  The larger the
amplitude of the error the lower the fidelity of the gate, as
Fig.~\ref{fig-4}(b) shows.

As mentioned before, the scheme is insensitive to temperature. If
the commensurability condition (\ref{cond}) is not perfectly
satisfied due to, for example, errors in timing of laser pulses,
or misalignment of the lasers, then the corresponding contribution
to the gate error is
\[
E=(C_1^4+C_2^4+C_1 C_2-6)/8 \] with
\begin{eqnarray} C_1&=&\exp
\left[- (1/2+ k_bT /\hbar\nu_c) |2\eta_c C_c|^2 \right] \\
C_2&=&\exp \left[- (1/2+ k_bT /\hbar\nu_r) |\eta_r C_r|^2 \right],
\end{eqnarray}
which is a smooth function of temperature $T$.

Finally, we would like to make some remarks regarding the
experimental implementation of this scheme. First, it is not
necessary to kick the atoms using pairs of counter-propagating
laser beams. The same effect (i.e.  a change of sign in $\eta$)
may also be achieved in current experiments by reverting the
internal state of both ions simultaneously. One then only needs a
laser beam aligned with the trap to kick the atoms, and another
laser orthogonal to the axis of the trap to produce the NOT-gate.

The second and more important remark is that it is possible to
avoid errors in the laser pulses by using more sophisticated
kicking methods. One possibility consists in using STIRAP
\cite{STIRAP-a,STIRAP-b}. Only one of the qubit states would be
connected by two on--resonance laser beams to a third atomic
state, $\ket{e'}$. In the first part of the kicking process, the
Rabi frequencies of the lasers $\Omega_a$ and $\Omega_b$ are
adiabatically switched on an off respectively. The momenta of both
laser beams should be different, so that as we slowly proceed from
$\Omega_a/\Omega_b\simeq 0$ to the opposite regime
$\Omega_b/\Omega_a \simeq 0$, the ions in the state $\ket{1}$ are
completely transferred to the new dark state $\ket{e'}$ and get a
kick $\ket{1}\to e^{i(\vec{k}_a-\vec{k}_b)\vec{x}}\ket{e'}$. Next
we must change the sense of the laser beams
($\vec{k}_{a,b}\to-\vec{k}_{a,b}$), and perform the adiabatic
transfer from $\Omega_b/\Omega_a \simeq 0$ to
$\Omega_a/\Omega_b\simeq 0$. The total transformation will be
$\ket{0}\to\ket{0}$, $\ket{1}\to
e^{2i(\vec{k}_a-\vec{k}_b)\vec{x}}\ket{1}$. The advantages of this
method are: (i) the system remains all the time in a dark state,
avoiding spontaneous emission; (ii) the process is insensitive to
fluctuations of the intensity; (iii) the duration of the pulse
need not be precisely adjusted, and (iv) the intensity of the
laser need not be the same for both ions.

Summing up, in this work we have developed a new concept of
two-qubit quantum gate for trapped ions, in which the trap
frequency poses no longer a limitation on the speed of the gate.
Rather than performing sideband transitions which weakly couple
the internal and the motional states of the atoms, we suggest to
push the atoms resonantly during very short times and along
different directions. The limitations in that case come from: (i)
the anharmonicities of the restoring force that the ions
experience when pushed far away from each other, and (ii) the
ability to control the laser pulses. The first limitation still
allows to perform the gates in a time which is three orders of
magnitude smaller than the one imposed by the trap frequency. The
second one can be overcome by using adiabatic passage techniques
which make these laser excitations tolerant against laser
imperfections. In any case, the rapid experimental progress in
laser control with very short pulses indicates that it may be
possible soon to perform quantum gates with a very high speed.  In
addition, our scheme is independent of the temperature, requires
no addressability, and works beyond the Lamb-Dicke regime.

J.I.C. thanks D. Leibfried for discussions and NIST for the
hospitality during his stay in Boulder. P.Z. thanks the Innsbruck
ion trap group for discussions. Part of this work was supported by
the EU IST project RESQ, the project TOPQIP, the DFG
(Schwer\-punkt\-programm Quanten\-informations\-verarbeitung) and
the Kompetenz\-netz\-werk Quanten\-informations\-verarbeitung der
Bayerischen Staatsregierung. Research at the University of
Innsbruck is supported by the Austrian Science Foundation,  EU
Networks and the Institute for Quantum Information.


\begin{thebibliography}{99}

\bibitem{PhysicsTodayMay2003}{B. G. Levi, Physics Today (2003).}

\bibitem{CiracZoller1}{J. I. Cirac, and P. Zoller, Phys. Rev. Lett.
    \textbf{74}, 4091 (1995).}

\bibitem{Wineland95}{C. Monroe \textit{et al.}, Phys.  Rev.  Lett.
    \textbf{75}, 4714 (1995).}

\bibitem{Wineland1}{B. DeMarco \textit{et al.}, Phys. Rev. Lett.
    \textbf{89}, 267901 (2002).}

\bibitem{Wineland2}{D. Leibfried \textit{et al.}, Nature \textbf{422}, 412
    (2003).}

\bibitem{Blatt}{F. Schmidt--Kaler \textit{et al.}, Nature \textbf{422},
    408 (2003).}

\bibitem{Wineland3}{Q. A. Turchette \textit{et al.}, Phys. Rev.  Lett.
    \textbf{81}, 3631 (1998).}

\bibitem{Wineland4}{V. Meyer \textit{et al.}, Phys. Rev. Lett.
    \textbf{86}, 5870 (2001).}

\bibitem{Wineland5}{C. A. Sackett \textit{et al.}, Nature \textbf{404},
    256 (2000).}

\bibitem{Scalable-a}{J. I. Cirac, and P. Zoller, Nature \textbf{404}, 579
    (2000).}

\bibitem{Scalable-b}{D. Kielpinksi, C. Monroe, and D. J. Wineland, Nature
    \textbf{417}, 709 (2002).}

\bibitem{papermotionWineland}{M. A. Rowe, \textit{et al.}, Quantum
    Inf. and Comp. \textbf{2}, 257 (2002).}

\bibitem{Proposals-b}{A. S{\o}rensen and K. M{\o}lmer, Phys. Rev. A
    \textbf{62}, 022311 (2000).}

\bibitem{Proposals-c}{A. S{\o}rensen, and K. M{\o}lmer, Phys. Rev.
    Lett. \textbf{82}, 1971--1974 (1999).}

\bibitem{Proposals-d}{D. Jonathan, M. B. Plenio, and P. L. Knight,
    Phys. Rev. A \textbf{62}, 042307 (2000).}

\bibitem{ProposalsMilburn}{G. J. Milburn, S. Schneider, and D. F. V.
    James, Fort. Phys. \textbf{48}, 801--810 (2000).}

\bibitem{Proposals-a}{A notable exception is J. F. Poyatos., J. I.
    Cirac, and P.  Zoller, Phys. Rev.  Lett. \textbf{81}, 1322 (1998).
    However, here highly non--harmonic traps are required and still
    the gate time is limited by the trap frequency.}

\bibitem{Nielsen}{M. A. Nielsen, and I. L. Chuang, \textit{Quantum
      Computation and Quantum information} (Cambridge Univ.  Press,
    Cambridge, 2000).}

\bibitem{STIRAP-a}{M. Weitz, B. C. Young, and S. Chu, Phys. Rev. A
    \textbf{50}, 2438 (1993).}

\bibitem{STIRAP-b}{K. Bergmann, H. Theuer, and B. W. Shore, Rev. Mod.
    Phys. \textbf{70}, 1003 (1998).}

\end{thebibliography}
\end{document}